\RequirePackage{fix-cm}
\documentclass[twocolumn]{svjour3}          
\smartqed  

\usepackage[square]{natbib}

\usepackage{graphicx}
\usepackage{siunitx}
\usepackage[draft]{listofsymbols}
\usepackage{amsmath,amssymb}
\usepackage{shellesc}
\usepackage{tikz}
\usetikzlibrary{external}
\usepackage{pgfplots} 
\usepgfplotslibrary{external}
\usepackage{pgfgantt}
\usepackage{pdflscape}
\usepackage[export]{adjustbox}
\usepackage[permil]{overpic}
\usepackage{subfig}
\pgfplotsset{compat=newest} 
\pgfplotsset{plot coordinates/math parser=false}
\usepackage{verbatim} 
\usepackage{ marvosym }
\usepackage[normalem]{ulem}
\usepackage{dblfloatfix}    
\usepackage{multirow}

\usetikzlibrary{shapes.arrows}

\usepackage[colorlinks,linkcolor=blue,citecolor=blue,urlcolor=blue,bookmarks=false,hypertexnames=true]{hyperref}

\usepackage[switch,columnwise]{lineno}

\usepackage{dblfloatfix}

\newcommand{\uproman}[1]{\uppercase\expandafter{\romannumeral#1}}

\begin{document}

\sloppy

\title{Species Transport Driven by Droplet Impact in Wavy Thin Films}

\author{Hatim Ennayar \and Frederik Roy Patria \and Jeanette Hussong
}
 
 \institute{
 {Hatim Ennayar} (\Letter)  \and Frederic Roy Patria \and Jeanette Hussong \at
             Institute for Fluid Mechanics and Aerodynamics\\
             TU Darmstadt 
             \\ Flughafenstr. 19, 64347, Darmstadt-Griesheim \\ 
              \email{ennayar@sla.tu-darmstadt.de}            
                               }

\date{Received: date / Accepted: date}

\maketitle

\begin{abstract}
Droplet impact on thin liquid films is commonly studied on quiescent surfaces, although practical systems often involve residual capillary waves generated by preceding droplets. This study examines how such traveling waves modify impact dynamics and mixing. Controlled surface disturbances were produced using an acoustic excitation system that replicated droplet-induced waves, and a two-color laser-induced fluorescence method was implemented to obtain simultaneous measurements of film thickness and dye concentration. Impacts on wavy films deviated markedly from quiescent conditions. Rim evolution, cavity collapse, and jet formation became asymmetric, governed by the phase of the wave relative to the impact. These behaviors were linked to local variations in film depth, which redirected cavity retraction and the associated mixing flow. Reconstructed concentration fields confirmed that droplet liquid is displaced according to these depth gradients, producing asymmetric mixing at moderate Weber numbers. A dimensionless asymmetry index quantified the dependence on wave amplitude, phase, and distance from the acoustic wave generator. At higher Weber numbers, inertial mixing attenuated these effects, and the dynamics approached those of static films.

\end{abstract}

\section{Introduction}
\label{sec:1}
Droplet impact onto liquid films is a classic fluid dynamics problem with broad relevance, from inkjet printing (\citealt{ikegawa2004droplet,mu2017inkjet}) to rainfall (\citealt{choo2018resistance,yu2022force}), spray cooling (\citealt{gajevic2023spray,zhao2022dynamics}), and spray coating (\citealt{butt2022thin,huang2018understanding}). Extensive research over the past decades has elucidated impact phenomena on quiescent surfaces, including deep pools (\citealt{ray2015regimes,castillo2015droplet,minami2022cavity,agbaglah2015drop,saha2019kinematics}) and thin stationary films (\citealt{rajendran2023predicting,parmentier2023drop,tang2019spreading,bernard2021miscibility}). When a droplet impacts a thin liquid film, it typically forms a crater and an outward-propagating rim, which may subsequently develop a crown and potentially followed by a vertical jet and secondary droplets if splashing occurs. The dynamics of this process are primarily governed by the Reynolds number, $Re = \frac{UD}{\nu}$, the Weber number, $We = \frac{\rho DU^2}{\sigma}$ and the dimensionless film thickness $\delta = \frac{h}{D}$, which relates the film thickness $h$ to the droplet diameter $D$. Here, $U$ is the droplet impact velocity, $\nu$ is the kinematic viscosity, $\rho$ is the liquid density, and $\sigma$ is the surface tension. These parameters have been extensively reviewed in recent literature (\citealt{yarin2006drop,josserand2016drop,breitenbach2018drop}), offering detailed insights into their influence on droplet impact outcomes.

In realistic industrial and natural scenarios, the film surface is rarely still. Prior droplets and ambient disturbances often generate traveling surface waves that persist until the next droplet arrives. Although most fundamental studies focus on quiescent films, recent investigations have started addressing the complexity of dynamic surfaces. Impacts on flowing or wavy films have been shown to alter cavity morphology, splashing behavior and mixing efficiency. In particular, droplet impacts on flowing films exhibit distinct behaviors due to streamwise velocity gradients and surface deformations that produce effective wave-like inclinations. These kinematic asymmetries lead to tilted jets (\citealt{alghoul2011normal}), asymmetric crowns (\citealt{adebayo2017droplet,che2015impact}), and shifted splashing thresholds (\citealt{gao2015impact,castrejon2016droplet}), fundamentally altering impact outcomes compared to static films. In contrast, other studies focus on pre-existing capillary waves superimposed on nominally quiescent films (\citealt{van2015single,khan2019experimental}). For instance, waves generated by lateral vibrations were investigated by \citealt{khan2020droplet}. Such surface waves, although small in amplitude, can influence the dynamics of the impact and mixing. Beyond thin films, the influence of capillary waves has been studied in deep pool configurations. Both \citealt{gupta2020splashing} and \citealt{singh2022droplet} showed that planar capillary waves generate unequal crests and troughs around the cavity, producing bent jets and directional secondary droplet ejection due to asymmetric collapse. \citealt{singh2022droplet} further highlighted that crest versus trough impacts lead to fundamentally different jet morphologies, where crest impacts yield deeper craters and slower, thicker Worthington jets, while trough impacts favor slender, inertia-dominated jets that fragment quickly.

Yet despite these advances, to the authors’ knowledge, no prior study has investigated droplet impacts on traveling capillary waves with properties analogous to those generated by a preceding droplet, especially in the thin film regime. The present work addresses this gap by experimentally isolating wave effects through a controlled wave generator and quantifying their influence on mixing.

To quantify the mixing dynamics under these wave-perturbed conditions, two-color laser-induced fluorescence (\citealt{coppeta1998dual,sakakibara2004measurement}) (2C-LIF) technique is employed. Conventional single-color Laser-induced fluorescence (LIF) is a well-established optical technique for measuring scalar quantities such as concentration, temperature, and film thickness, and has been successfully applied across a broad range of fluid mechanics problems (\citealt{seitzman1985instantaneous,vasudevan2018laser,yang2025simultaneous}). The authors previously used this technique to investigate droplet mixing dynamics on quiescent thin liquid films (\citealt{ennayar2023lif}). However, that implementation required separate experiments to determine film thickness and dye concentration, thereby increasing the experimental effort and introducing additional sources of systematic uncertainty. To overcome this limitation, the present study builds on the 2C-LIF methodology developed in our previous work (\citealt{ennayar2026two}). The method enables the simultaneous measurement of local film thickness and dye concentration from a single image acquisition, thereby avoiding the sequential measurements required in conventional single-color LIF. Over the past decades, 2C-LIF has proven to be a powerful diagnostic tool in fluid mixing studies (\citealt{koegl2020novel,zhang2014simultaneous,mohri2011imaging}). For instance, \citealt{kong2018dual} used 2C-LIF to spatially resolve concentration maps in the wake of rising bubbles by converting pH-field measurements into quantitative dissolved CO$_2$ concentration fields. More recently, \citealt{ulrich2024two} applied 2C-LIF to simultaneously capture temperature and concentration fields in binary-mixture droplets during jet breakup.

In what follows, the influence of surface waves on the spatiotemporal mixing dynamics resulting from droplet impact onto wavy thin films is investigated. Quantitative concentration fields are extracted using the 2C-LIF technique, allowing the effects of wave-induced topography on mixing to be characterized and the conditions under which these effects become negligible to be identified.

\section{Materials and methods}
\label{sec:2}
In the following, the experimental setup, sample preparation and calibration procedure are explained in detail. Here a focus is put on the realization of 2C-LIF and capillary-wave generation.

\subsection{Experimental setup}
\label{sec:2.1}
\begin{figure}
	\centering
	\includegraphics[scale=0.75]{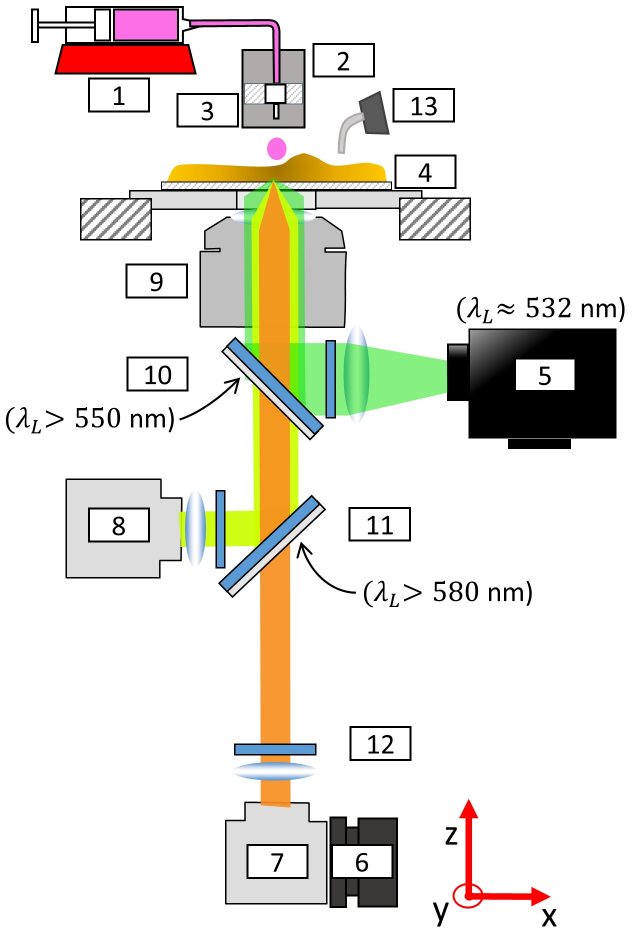}
	\caption{
		Sketch of setup: (1) Syringe pump, (2) z-Traverse, (3) Cannula, (4) Wavy thin liquid film on FTO glass substrate, (5) High power LED, (6) x,y,z-Traverse, (7) HS-Camera, (8) HS-Camera, (9) Microscope Objective, (10,11,12) Dichroic mirror with bandpass filters, (13) Acoustic wave generator.
	}
	\label{fig:1}
\end{figure}

The experimental configuration was based on a previously established arrangement developed for two-color LIF investigations of droplet impact on thin liquid films (\citealt{ennayar2026two}). this approach enables the simultaneous quantification of local film thickness and tracer concentration. A schematic overview of the setup is provided in Fig. \ref{fig:1}.

Water droplets were generated using a blunt-tip needle (Braun GmbH) connected via a tube to a $\SI{5}{\milli\litre}$ syringe (Braun GmbH) driven by a syringe pump (Aladdin AL-1010, WPI). The needle was mounted on a computer-controlled traverse, which allowed precise adjustment of the release height and thus control of the impact velocity. The droplet diameter was maintained at $D = 2.225 \pm 0.025~\mathrm{mm}$. Impacts occurred on a borosilicate glass substrate of $50 \times 50~\mathrm{mm^2}$ (Sigma-Aldrich) that was cleaned in an ultrasonic bath with isopropanol and rinsed with deionized water before each run. A thin water film was then spread on the substrate, with the contact line pinned along the four edges to ensure repeatable initial conditions. Film thicknesses of $h = 500~\mathrm{\mu m}$ and $800~\mathrm{\mu m}$ were investigated. The thickness before impact was monitored by a chromatic-confocal point sensor (confocalDT IFS2407-0.8, Micro-Epsilon) providing an accuracy of $\pm 0.4~\mathrm{\mu m}$. The corresponding impact conditions covered Reynolds numbers $Re = 1800$, $3000$, and $4200$ and Weber numbers $We = 24$, $54$, and $110$, respectively.

To examine the interaction between the impacting droplet and a wavy liquid film, a custom-built acoustic wave generator was employed to excite controlled capillary waves on the film surface. The device comprised a compact $45~\mathrm{mm}$ speaker (Visaton BF45). The acoustic field emitted by the speaker was focused using a stainless-steel funnel and subsequently guided through a $2.5~\mathrm{mm}$-diameter tube. This narrowing ensured that the effective emission area is of the same order of magnitude as the droplet size. The outlet tube was bent by approximately $45^{\circ}$, allowing the acoustic wave to impinge vertically on the liquid surface while the generator remained positioned off-axis. This configuration avoided obstruction for the falling droplet. The speaker, funnel, and tube were connected via a 3D-printed plastic mount which was fixed to a motorized arm that allowed automated adjustment of the generator position in three directions. This mechanism enabled precise control of the distance between the wave source and the droplet impact location. The acoustic excitation offered several advantages over mechanical or piezoelectric actuation methods. It provided contactless generation of capillary waves, eliminating contamination risks. Furthermore, the wave amplitude and frequency could be readily tuned by varying the electrical driving parameters, allowing precise control of the wave characteristics. A photodiode-based droplet detection unit was integrated below the needle, triggering the wave generator to synchronize droplet release with the wave excitation. This allowed systematic variation of the phase relationship between the impinging droplet and the traveling surface wave.

The optical arrangement consisted of a custom-built microscope designed to enable high-speed, coaxial fluorescence imaging of the droplet impact region. The entire 2C-LIF assembly was mounted on a precision three-axis traverse with a minimum step size of $1.25~\mathrm{\mu m}$, allowing fine adjustment of the optical position relative to the impact area for image acquisition. Illumination was provided by a continuous-wave high-power green LED ($\lambda \approx 532~\mathrm{nm}$, ILA iLA.LPS v3), which was coupled coaxially into the optical path. The emitted fluorescence was collected through a long-working-distance $1\times$ microscope objective (Infinity Photo-Optical IF-3) and spectrally separated into two detection bands centered at $561 \pm 10.5~\mathrm{nm}$ and $624 \pm 23~\mathrm{nm}$ using dichroic mirrors and band-pass filters (Thorlabs DFM1/M). Two synchronized high-speed CMOS cameras, a Phantom T3610 (12-bit, $1200 \times 800$ pixels) and a Phantom T1340 (12-bit, $2048 \times 1952$ pixels), simultaneously recorded the fluorescence signals at $6000~\mathrm{fps}$. The intensities were later combined into their sum and ratio, providing two independent observables that form the basis for determining the local film thickness and dye concentration. Rhodamine~B (RhB, Carl Roth) and Rhodamine~6G (Rh6G, Carl Roth) were employed as tracers. Rhodamine~B was added at $7.5 \times 10^{-5}\mathrm{M}$ to both droplet and film, while Rhodamine~6G was introduced only into the film at the same concentration.

In addition to the bottom-view 2C-LIF imaging, side-view shadowgraphy was conducted to visualize the overall impact dynamics and the influence of pre-existing capillary waves. The shadowgraphy system consisted of the same Phantom T1340 high-speed camera operated at $6000~\mathrm{fps}$ equipped with a Nikkon AF Micro 60 mm lens. The scene was back-illuminated by an LED (Veritas Constellation 120E). With the experimental configuration established, the following section presents the calibration procedure for the 2C-LIF measurements and describes the methods used to characterize the acoustic wave generator and determine the phase relation between the droplet impact and the surface wave.

\subsection{2C-LIF Method}
\label{sec:2.2}

The local film thickness and dye concentration were obtained using the two-color laser-induced fluorescence (2C-LIF) method developed in our previous work (\citealt{ennayar2026two}). Only the main aspects relevant to the present experiments are summarized here.

The fluorescence signal depends on the excitation intensity, fluorescence quantum yield, molar absorptivity, optical path length, and dye concentration. In simplified form, it can be written as (\citealt{guilbault1973})
\begin{equation}
	I_{f} = A I_{0} \Phi \epsilon L C ,
	\label{eq:1}
\end{equation}
where $L$ corresponds to the local film thickness and $C$ is the dye concentration. Since the present experiments were conducted outside the strictly linear fluorescence regime, a multidimensional calibration was required to relate the measured fluorescence intensities to film thickness and concentration. Fluorescence was recorded simultaneously in two spectral channels, yielding the intensities $I_{1}$ and $I_{2}$. These signals were combined into the total intensity $S = I_{1}+I_{2}$ and the intensity ratio $R = \frac{I_{2}}{I_{1}}$. The calibration model then mapped the measured quantities $S(x,y)$ and $R(x,y)$ to the local film thickness $h(x,y)$ and dye concentration $C(x,y)$.

The calibration was performed on the same substrate and with the same optical alignment as used in the wave experiments. Films of known thickness between $0$ and $\SI{1000}{\micro\meter}$ were prepared, while the dye concentrations were varied over the relevant experimental range. For each calibration condition, synchronized image pairs were acquired, geometrically aligned onto a common grid, filtered, and processed pixel-wise to account for spatial variations in illumination and optical collection efficiency. Details of the calibration procedure, image registration, regression model, and uncertainty analysis are given in (\citealt{ennayar2026two}). The calibration accuracy was verified using independent test measurements. The resulting mean absolute deviation was $\SI{24.2}{\micro\meter}$ for film thickness and $1.6\times10^{-6}$ M for concentration. These values confirm that the method is sufficiently accurate for simultaneous reconstruction of film thickness and dye concentration in the present wave experiments.

\begin{figure}
	\centering
	\includegraphics[scale=0.9]{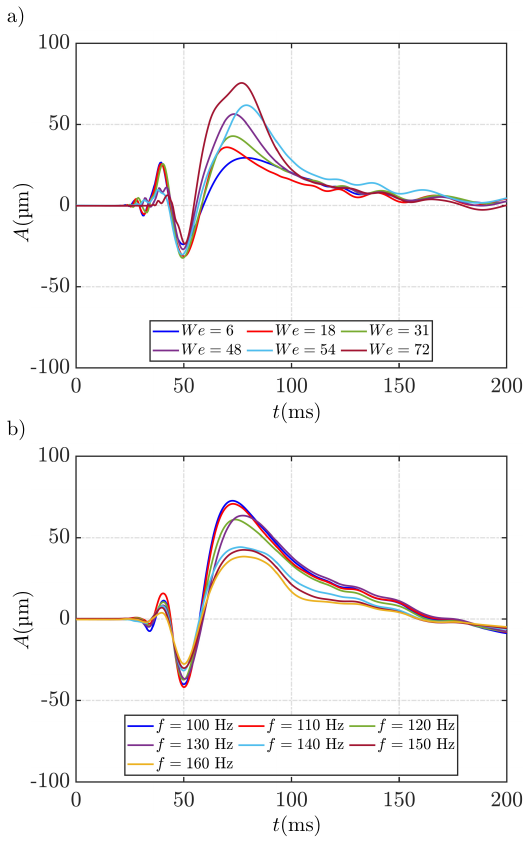}
	\caption{Time-resolved amplitude $A(t)$ of capillary waves measured at $d_s = 10~\mathrm{mm}$ from the center of the wave excitation. \textbf{a)} Wave generated by droplet impacts on a thin film of thickness $\delta = 0.22$ under varying Weber numbers. \textbf{b)} Acoustically-induced waves at different frequencies $f$.}
	\label{fig:3}
\end{figure}

Compared with single-color LIF, the 2C-LIF approach provides two independent observables from one image sequence. Therefore, film thickness and dye concentration can be determined simultaneously, avoiding the sequential measurements required in conventional single-color LIF (\citealt{ennayar2023lif}).

\begin{figure*}[b]
	\centering
	\includegraphics[scale=0.7]{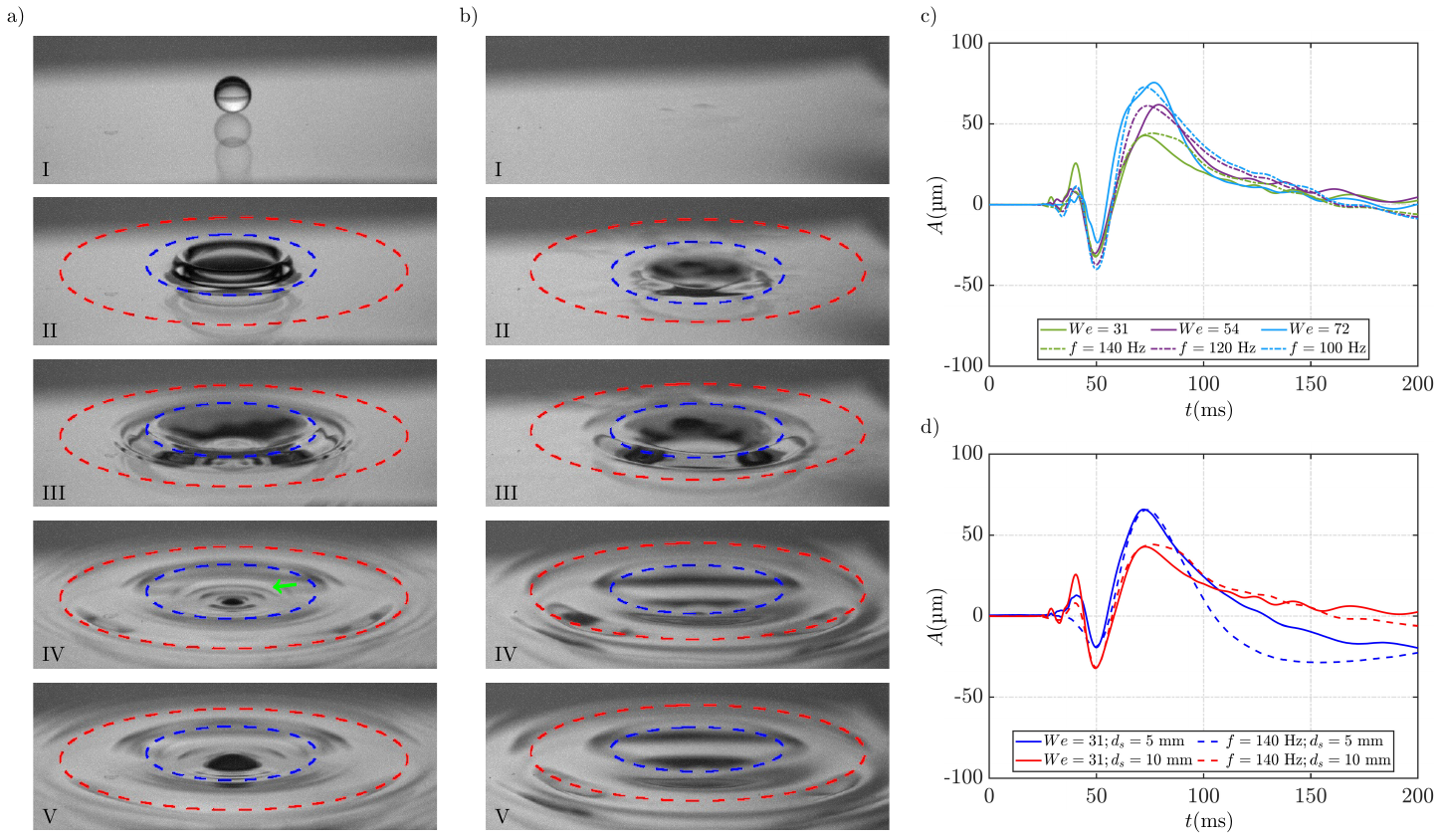}
	\caption{Oblique-view image sequences illustrating wave evolution following \textbf{a)} droplet impact and \textbf{b)} acoustic excitation on a thin film of thickness $\delta = 0.22$, corresponding to $We = 31$ and $f = \SI{140}{\hertz}$, respectively. Dashed blue and red ellipses mark radial positions $d_s = 5~\mathrm{mm}$ and $d_s = 10~\mathrm{mm}$, where waves' amplitudes were measured. The green arrow points to the small secondary ripples on the surface. The images I–V represent times of $0$, $5$, $10$, $14$, and $17~\mathrm{ms}$. \textbf{c)} Time evolution of wave amplitude at $d_s = 10~\mathrm{mm}$ comparing droplet-induced waves and acoustically-induced waves. \textbf{d)} Comparison of wave amplitude profiles for $We = 31$ and $f = \SI{140}{\hertz}$ at two measurement distances $d_s = 5~\mathrm{mm}$ (blue) and $d_s =10~\mathrm{mm}$ (red).}
	\label{fig:4}
\end{figure*}

\subsection{Capillary wave generation}
\label{sec:2.3}
A validation of the acoustic wave generator was first performed by comparing the capillary waves it produced with those generated naturally by droplet impact. The objective was to create controlled surface perturbations resembling droplet-induced waves, but with well-defined and reproducible characteristics. This approach enables a first systematic study of how such surface waves influence subsequent droplet impacts in environments where multiple droplets interact with the same liquid film. By decoupling the surface topography effect from solute transport induced by earlier droplets, the intrinsic influence of the wavy interface can be investigated under controlled and repeatable conditions.

The surface wave amplitude $A(t)$ was measured using the chromatic-confocal point sensor positioned at a distance $d_s = 10~\mathrm{mm}$ from the center of the wave excitation. The temporal evolution of the film surface was evaluated from the variation of the measured distance signal relative to the quiescent film level. Fig. \ref{fig:3}a shows the amplitude evolution for waves generated by droplet impact on a film of dimensionless thickness $\delta = 0.22$ at different impact conditions. The corresponding Reynolds and Weber numbers are summarized in Table \ref{tab:1}. These conditions were varied by adjusting the droplet release height, thereby changing the impact velocity. Three positive peaks were observed with the largest amplitude strongly depending on the impact velocity. Increasing the impact energy resulted in a pronounced increase in the maximum wave amplitude.

Acoustically induced surface waves were generated by applying a a short sinusoidal burst of $20~\mathrm{ms}$ at defined driving frequencies. Fig. \ref{fig:3}b presents the temporal evolution of the amplitude measured for the same film thickness ($\delta = 0.22$) at excitation frequencies between $\SI{100}{\hertz}$ and $\SI{160}{\hertz}$. The general waveform closely resembled that of droplet-induced surface waves. Moreover, the maximum amplitude decreased with increasing driving frequency.This inverse relationship can be attributed to the mechanical response of the loudspeaker diaphragm. At lower frequencies, the cone excursion required to produce a given sound pressure level is larger (\citealt{small1972direct,kinsler2000fundamentals}), resulting in higher air-volume displacement and consequently stronger acoustic loading on the liquid surface and thus the resulting higher surface wave amplitude.

\begin{table}
\centering
\begin{tabular}{|c||c|}\hline
Weber number $We$ & Reynolds number $Re$ \\\hline 
6 & 979 \\\hline
18 & 1793 \\\hline
31 & 2225 \\\hline
48 & 2781 \\\hline
54 & 3000 \\\hline
72 & 3400 \\\hline
\end{tabular}
\caption{Impact conditions for droplet-induced surface waves shown in Fig.~\ref{fig:3}a.}
\label{tab:1}
\end{table}

The preceding characterization of the acoustically generated surface waves confirmed their tunability through the driving frequency. To verify that they reproduced the same surface dynamics as droplet-induced waves, both cases were examined and compared as shown in Fig. \ref{fig:4}c. In this comparison, the time-resolved amplitudes of both droplet-induced and acoustically generated waves were plotted together, measured at $d_s= 10~\mathrm{mm}$ from the excitation center on a film of thickness $\delta = 0.22$. The dataset includes three droplet impact conditions ($We = 31$, $54$, and $72$) and three excitation frequencies ($f = 100$, $120$, and $\SI{140}{\hertz}$). The resulting curves exhibit nearly identical temporal evolution, confirming that the two excitation mechanisms produce comparable surface response. Decreasing the excitation frequency resulted in larger wave amplitude, reproducing the same trend observed when increasing the droplet impact velocity. This demonstrate that tuning the acoustic frequency provides a reliable means of emulating the hydrodynamic response associated with different droplet impact conditions. 

Beyond this quantitative agreement, the surface deformation was also visualized using inclined-view imaging to compare the spatiotemporal development of both cases (Figs. \ref{fig:4}a,b). The selected cases correspond to $We=31$ for the droplet-induced waves and $f=\SI{140}{\hertz}$ for the acoustically generated ones, both at $\delta = 0.22$. The sequence of images corresponds to times $t = 0$, $5$, $10$, $14$, and $17~\mathrm{ms}$, capturing the early crater formation and subsequent propagation of surface waves. It should be noted that the time origin in the visualizations ($t = 0~\mathrm{ms}$) corresponds to the moment when the droplet or the sound wave first interacted with the liquid surface, while in the confocal measurements the temporal evolution is shown relative to the start of the sensor recording. Shortly after droplet impact or acoustic excitation ($t = 5~\mathrm{ms}$), a crater of comparable diameter was observed in both cases. The main distinction lies in the rim height, where the droplet-induced deformation produced a significantly higher rim due to the additional inertial contribution of the impacting mass. At $t = 10~\mathrm{ms}$, the radial propagation of the waves became apparent, exhibiting similar wavelength and propagation speed for both excitation types. The droplet-induced case, however, displayed fine capillary ripples ahead of the main wavefront. These small-amplitude ripples can also be observed in Fig. \ref{fig:3}a at the onset of the amplitude evolution, whereas they are absent in the acoustically generated case shown in Fig. \ref{fig:3}b.

By $t = 14~\mathrm{ms}$, a jet formation was observed in the droplet case, which was absent in the acoustic one. This jetting led to the generation of small secondary ripples on the surface (highlighted by the green arrow in Fig. \ref{fig:4}a), which later manifested as periodic oscillations in the temporal amplitude evolution.

The dashed blue and red contours in Figs.~\ref{fig:4}a,b mark the radial positions at which the confocal sensor recorded the film height during the experiments, corresponding to $d_s = 5~\mathrm{mm}$ and $d_s = 10~\mathrm{mm}$, respectively. The resulting amplitude evolution at these locations are plotted in Fig. \ref{fig:4}d. When the measurement location was shifted from $d_s = 10~\mathrm{mm}$ to $d_s = 5~\mathrm{mm}$, the initial phase of the wave evolution remained similar, but the overall amplitude increased as a result of the reduced distance to the excitation point. In this closer position, the amplitude exhibited a distinct drop into negative values after the main peak. This behavior results from the confocal sensor capturing the retraction of the crater as the liquid surface relaxed toward its equilibrium level.

\begin{figure}[b]
	\centering
	\includegraphics[scale=0.45]{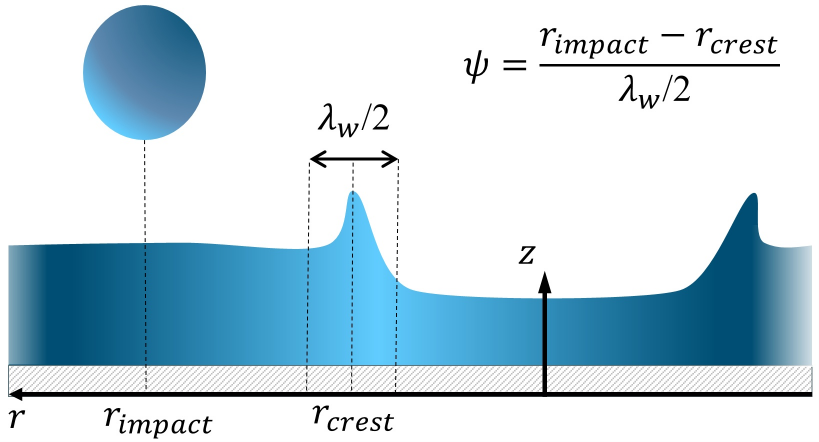}
	\caption{Schematic of droplet impact on wavy thin film. The phase parameter $\psi$ quantifies the impact position relative to the wave crest.}
	\label{fig:6}
\end{figure}

\begin{figure*}[b]
	\centering
	\includegraphics[scale=0.7]{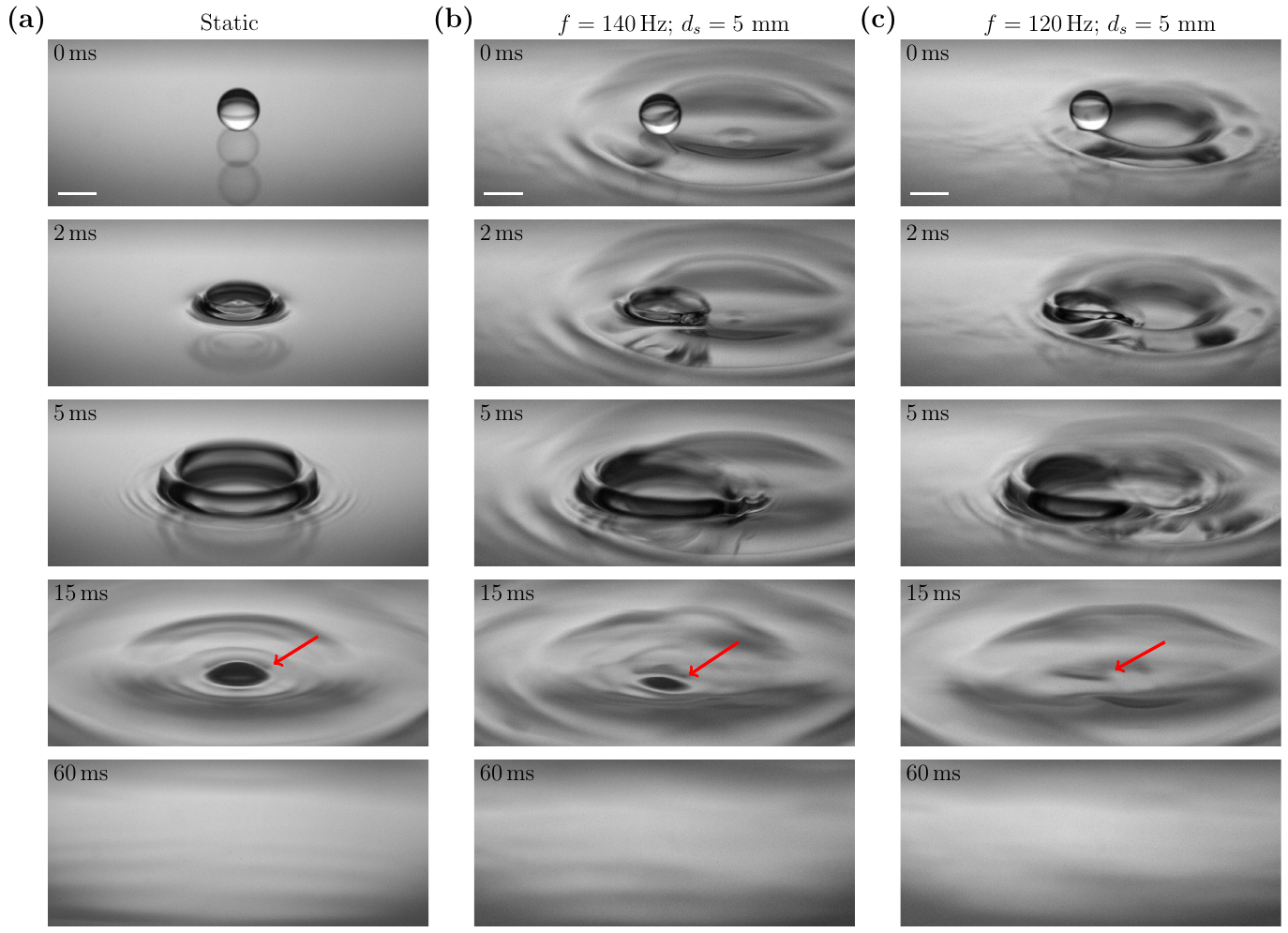}
	\caption{Oblique-view visualizations of droplet impact on a wavy liquid film for three acoustic excitation conditions at $\delta = 0.22$, $We = 54$, $Re = 3000$ and $\psi \approx -0.25$. \textbf{a)} $f = \SI{120}{\hertz}$ and $d_s = 5~\mathrm{mm}$. \textbf{b)} $f = \SI{140}{\hertz}$ and $d_s = 5~\mathrm{mm}$. \textbf{c)} $f = \SI{200}{\hertz}$ and $d_s = 7.5~\mathrm{mm}$. The time instants are $t = 0$, $2$, $5$, $15$, and $60~\mathrm{ms}$. Red arrows highlight the influence of surface waves on jet formation. Scale bar is equivalent to 2 mm.}
	\label{fig:7}
\end{figure*}

A further difference was that, for $d_s = 5~\mathrm{mm}$, the crater depth in the acoustic case exceeded that of the droplet-induced one, which can be attributed to the absence of an additional liquid mass (droplet) in the acoustic excitation. Furthermore, the small-amplitude ripples superimposed on the droplet-induced waveform, which were absent in the acoustic case for $d_s = 5~\mathrm{mm}$, correspond to the fine capillary disturbances triggered by jet formation and subsequent surface recoil, as mentioned earlier. These observations confirm that the acoustic excitation successfully reproduced surface-wave characteristics similar to those of the droplet-induced case, while offering a controllable and repeatable method for generating analogous surface topographies without introducing additional mass or concentration perturbations.



\section{Effect of surface waves on droplet impact dynamics}
\label{sec:3}
This section addresses the influence of pre-existing surface waves on droplet impact dynamics in thin liquid films. The local wave topography modifies the symmetry of the impact and thereby affects spreading, jet formation, film-thickness evolution, and species transport. To analyze these effects systematically, the relevant control parameters are first introduced, followed by an examination of the qualitative impact dynamics, the reconstructed film thickness, and the resulting mixing asymmetry.

\begin{figure*}[b]
	\centering
	\includegraphics[scale=0.7]{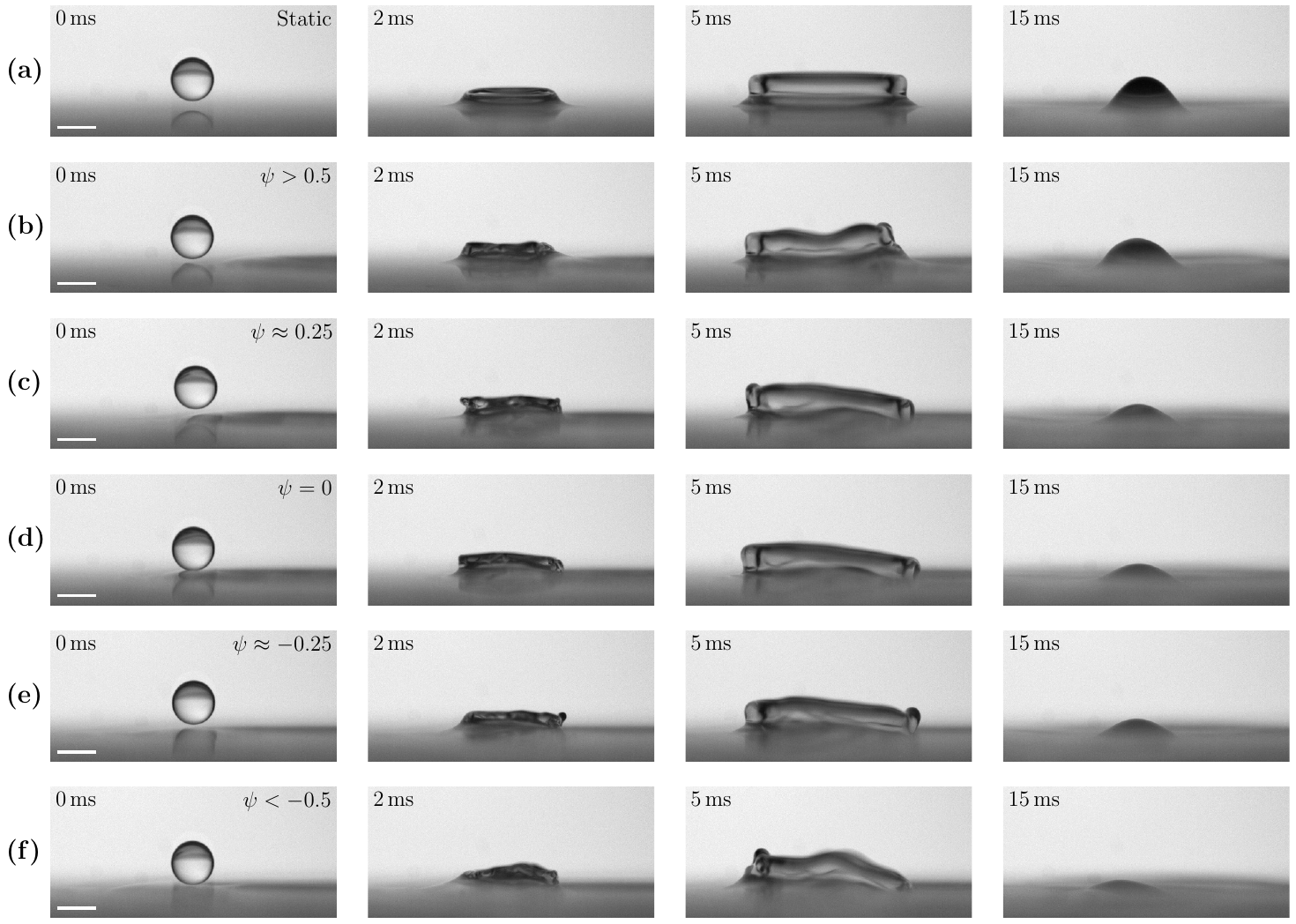}
	\caption{Side-view visualizations of droplet impact on a liquid film at $\delta = 0.36$, $We = 54$, and $Re = 3000$. \textbf{a)} static film. \textbf{b)} to \textbf{f)} wavy film at $f = \SI{120}{\hertz}$ and $d_s = 5~\mathrm{mm}$ for $\psi>0.5$, $\psi \approx 0.25$, $\psi=0$, $\psi \approx -0.25$ and $\psi< -0.5$, respectively. The time instants are $t = 0$, $2$, $5$ and $15~\mathrm{ms}$. Scale bar is equivalent to 2 mm.}
	\label{fig:8}
\end{figure*}

\subsection{Impact phase and control parameters}
\label{sec:3.1}
To characterize the droplet impact relative to the wavy liquid surface, a nondimensional phase parameter $\psi$ is introduced (Fig. \ref{fig:6}):
\begin{equation}
	\psi = \frac{r_{\text{impact}}-r_{\text{crest}}}{\lambda_w/2},
	\label{eq:2}
\end{equation}
where $r_{\text{impact}}$ denotes the radial coordinate of the droplet impact position, $r_{\text{crest}}$ is the position of the wave crest, and $\lambda_w$ represents the wavelength of the acoustically generated surface wave. A value of $\psi = 0$ corresponds to an impact occurring exactly at the crest, while positive and negative values indicates that the droplet impacts ahead or behind the crest, respectively.

Based on this definition, five representative impact configurations are identified. In the ``pre-front" case ($\psi > 0.5$), the droplet impacts ahead of the approaching wave, before the surface has risen. In the ``front-slope" case ($\psi \approx 0.25$), the impact occurs during the steep rising phase of the wave. The crest case ($\psi = 0$) corresponds to an impact directly on the wave peak. In the ``back-slope" case ($\psi \approx -0.25$), the droplet encounters the descending part of the wave, while in the ``far-back" case ($\psi < -0.5$), the impact takes place within the trough following the wave. The impact phase was controlled precisely using the photodiode-based droplet detection system described in Section \ref{sec:2}. The detection signal triggered the loudspeaker pulse, ensuring synchronization between the droplet release and the initiation of the surface wave. By introducing a defined delay between the detection and the wave generator pulse, and knowing the droplet velocity, the impact phase $\psi$ could be accurately adjusted. This approach enabled the realization of repeatable impact at well-defined wave phases with high precision. 

In addition to the impact phase, the wave amplitude represents a key control parameter. It can be adjusted by changing the excitation frequency or by varying the distance $d_s$ between the droplet impact position and the acoustic source. $d_s$ influences not only the amplitude but also the local wave profile at the impact location. As shown previously, when the impact location lies closer to the wave source, residual curvature associated with the crater dynamics alters the surface elevation onto which the droplet impacts. The following analysis therefore investigates the combined influence of wave amplitude and phase on the droplet impact dynamics.

\subsection{Qualitative impact dynamics and jet suppression}
\label{sec:3.2}
The presence of surface waves modifies the early-stage dynamics of droplet impact on thin liquid films. In particular, the interaction between the spreading lamella and the local wave topology can alter the symmetry of the rim and influence the formation of the Worthington jet. In the following, the qualitative impact dynamics are examined by varying two key parameters of the wavy film: the wave amplitude and the phase of the wave at the moment of impact.

\subsubsection{Influence of the wave amplitude}
\label{sec:3.2.1}
Fig. \ref{fig:7} presents oblique-view visualizations of droplet impact on a wavy liquid film for three acoustic excitation conditions at constant $\delta = 0.22$, $We = 54$, and $Re = 3000$. The droplet impact the back-slope region of the formed wave ($\psi \approx -0.25$). Figs. \ref{fig:7}a-c correspond static film and perturbed film with excitation frequencies of $f = \SI{140}{\hertz}$ and $f = \SI{120}{\hertz}$ with the wave generator placed at a distance $d_s = 5~\mathrm{mm}$ from the droplet impact. The corresponding time instants are $t = 0$, $2$, $5$, $15$, and $60~\mathrm{ms}$.

Immediately after impact on a static liquid film, the rim evolves in an axisymmetric manner, as shown in Fig.~\ref{fig:7}a. The circular rim forms uniformly around the impact center and expands radially without noticeable asymmetry during the early stages of spreading. When a surface wave is present, this symmetry becomes progressively disturbed. For the case with excitation frequency $f = \SI{140}{\hertz}$ (Fig. \ref{fig:7}b). the wave amplitude is moderate and only weakly modifies the rim evolution. A slight asymmetry develops as the droplet spreads over the wavy surface and the section of the rim encountering the cavity region of the wave collapses within $4~\mathrm{ms}$. This behavior arises from the superposition of the downward motion of the local wave surface and the upward inertial rise of the forming rim, effectively suppressing formation on that side. When the frequency is decreased to $f = \SI{120}{\hertz}$ (Fig. \ref{fig:7}c), the surface wave amplitude increases, resulting in a stronger change of the rim evolution. Its collapse is in the case quicker and starts around $2~\mathrm{ms}$, whereas the asymmetry becomes more pronounced as the droplet spreads over the wavy film surface.

A further notable difference appears at $t = 15~\mathrm{ms}$, where the jet developing during droplet impact becomes progressively weaker with increasing influence of the surface wave. As can be seen for the case $f = 120~\mathrm{Hz}$ and $d_s = 5~\mathrm{mm}$, jet formation is completely suppressed, as shown with the red arrow in Fig. \ref{fig:7}c. This suppression arises from the interference between the upward flow generated by the cavity collapse following droplet impact and the simultaneous inward retraction of the pre-existing cavity generated by the acoustic excitation. The resulting counteraction diminishes the vertical momentum necessary for jet formation and may entirely suppress it under large-amplitude waves. As the surface wave amplitude decreases, this interaction becomes weaker and the jet progressively recovers, eventually reaching heights comparable to those observed for the static-film case.

\begin{figure}
	\centering
	\includegraphics[scale=0.925]{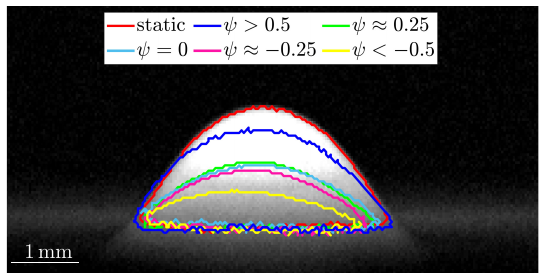}
	\caption{Jet height for different impact phases $\psi$, compared to the static case.}
	\label{fig:9}
\end{figure}

\begin{figure}[b]
	\centering
	\includegraphics[scale=0.9]{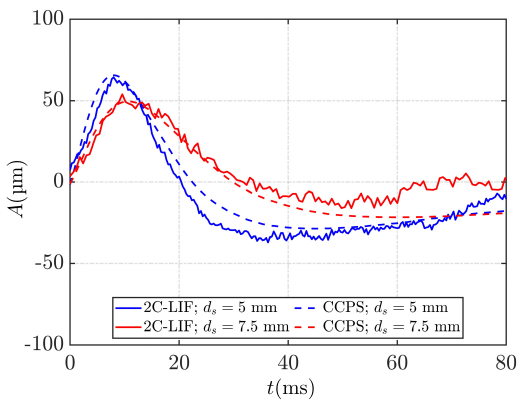}
	\caption{Temporal evolution of wave amplitude $A(t)$ measured at $d_s = 5~\mathrm{mm}$ (blue) and $d_s = 7.5~\mathrm{mm}$ (red) for $f = \SI{140}{\hertz}$ and $\delta = 0.22$. Solid lines indicate 2C-LIF thickness reconstructions; dashed lines show corresponding measurements from the chromatic-confocal point sensor (CCPS).}
	\label{fig:5}
\end{figure}

\subsubsection{Influence of the impact phase}
\label{sec:3.2.2}
Fig. \ref{fig:8} presents side-view visualizations of droplet impact at $\delta = 0.36$, $We = 110$, and $Re = 4200$ for six different surface conditions. Case (a) corresponds to impact on a quiescent film, whereas cases (b-f) show impacts on wavy films generated by an acoustic excitation at $f = \SI{120}{\hertz}$ for distinct impact phases: (b)~pre-front ($\psi > 0.5$), (c)~front-slope ($\psi \approx 0.25$), (d)~crest ($\psi = 0$), (e)~back-slope ($\psi \approx -0.25$) and (f)~far-back ($\psi < -0.5$). The corresponding times are $t = 0$, $2$, $5$, and $15~\mathrm{ms}$.

\begin{figure*}[b]
	\centering
	\includegraphics{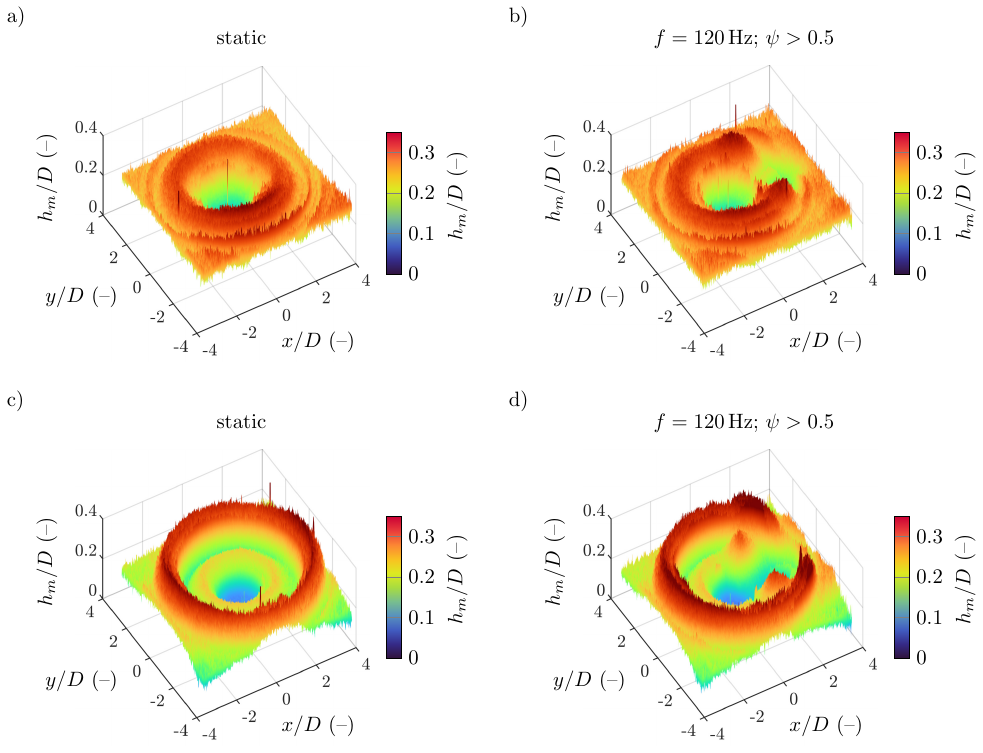}
	\caption{3D film height reconstruction for $\delta=0.22$ obtained using two-color LIF at $t = 10~\mathrm{ms}$. \textbf{a)} static film at $We = 54$ and $Re = 3000$. \textbf{b)} wavy film at $We = 54$, $Re = 3000$, $f= \SI{120}{\hertz}$, $d_s = 5~\mathrm{mm}$ and $\psi>0.5$. \textbf{c)} static film at $We = 110$ and $Re = 4200$. \textbf{d)} wavy film at $We = 110$, $Re = 4200$, $f= \SI{120}{\hertz}$, $d_s = 5~\mathrm{mm}$ and $\psi>0.5$.}
	\label{fig:10}
\end{figure*}

In the static film case (Fig. \ref{fig:8}a), the spreading rim develops into a pronounced corona as the lamella expands rapidly after impact. The upward lift is caused by the aerodynamic force on the spreading lamella as explained in the works of \citealt{riboux2014experiments} and \citealt{burzynski2020splashing}. At later times, the formation of a jet is observed. For the pre-front case shown in Fig. \ref{fig:8}b, the advancing spreading rim interacts with the incoming surface wave. The superposition of the upward-moving wavefront and the expanding liquid sheet leads to a localized elevation of the right-hand side of the corona at $t = 5~\mathrm{ms}$. The subsequent jet formation is slightly reduced in height compared with the static case. In the front-slope case (Fig.~\ref{fig:8}c), the influence of the surface wave becomes more evident. The liquid sheet rises asymmetrically, with the left portion of the crown tilted toward the direction of wave propagation. As the wave and the corona evolve, their superposition produced a local depression of the corona wall on the the right side, analogous to the deformation mechanisms observed previously for lower Weber number impacts illustrated in Fig.~\ref{fig:7}. Consequently, the resulting jet is shorter than in both the static and $\psi > 0.5$ cases. 

When the droplet impacts at the wave crest ($\psi = 0$, Fig.~\ref{fig:8}d), the right side of the rim subsides as the crown overlaps with the cavity formed by the pre-existing wave. The jet height remains low, comparable to the front-slope case. For the back-slope case ($\psi \approx -0.25$, Fig.~\ref{fig:8}e), the rim initially tilts in the opposite direction of the traveling wave, following the descending local interface profile. This behavior resembles droplet impact on an inclined wetted surface, as observed in the work of \citealt{chen2020experimental}, where the crown tilts in the direction of the inclination. The same is occurring in both $\psi \approx 0.25$ and $\psi \approx -0.25$ cases, with the crown is consistently leaning along the slope of the wavy interface at the impact location. The subsequent evolution again exhibits pronounced wave-sheet interaction, and the resulting jet height remains lower that for the quiescent reference. Finally, for the far-back case, the droplet impacts the trough of the wave. The right side of the corona exhibits the strongest depression, corresponding to the location of the cavity formed by the acoustic excitation. Furthermore, the rim geometry is significantly altered and the resulting jet is the lowest among all configurations illustrated in Fig. \ref{fig:9}.

\subsection{Thickness reconstruction on wavy films}
\label{sec:3.3}
To illustrate the accuracy of the 2C-LIF thickness reconstruction for dynamically varying films, the measured thickness evolution was compared with data obtained from the chromatic-confocal point sensor (CCPS). Fig.~\ref{fig:5} presents the temporal evolution of the wave amplitude at two radial positions, $d_s = 5~\mathrm{mm}$ and $d_s = 7.5~\mathrm{mm}$, for an excitation frequency of $f = 140~\mathrm{Hz}$ and film thickness $\delta = 0.22$. The curves correspond to the main surface wave, whose evolution is of particular interest for the subsequent droplet impact on wavy film experiments. The 2C-LIF thickness reconstruction reproduces the waveform captured by the CCPS with a typical deviation of approximately $\SI{5}{\micro\meter}$ at both measurement positions. The high-frequency noise level observed in the 2C-LIF measurements originates from pixel-scale intensity fluctuations in the fluorescence signal and the sensitivity of the thickness reconstruction to small variations in the recorded emission intensity corresponding to small thickness variations. The good agreement between the two measurement techniques further confirms the capability of the 2C-LIF method to resolve transient film thickness variations, consistent with the validation results presented in our previous work (\citealt{ennayar2026two}).

Representative contour plots derived from 2C LIF are shown in Figs. \ref{fig:10}a and b for $We=54$, as well as Figs. \ref{fig:10}c and d for $We=110$. For each Weber number, the capillary waves on a static film (a and c), and on a wavy film (b and d) are compared for $f = \SI{120}{\hertz}$ and $d_s = 5~\mathrm{mm}$. For the wavy film, an impact phase $\psi > 0.5$ is chosen. All thickness reconstructions are taken at $t = 10~\mathrm{ms}$, with the normalized film thickness plotted as $h_m/D$.


\begin{figure}
	\centering
	\includegraphics{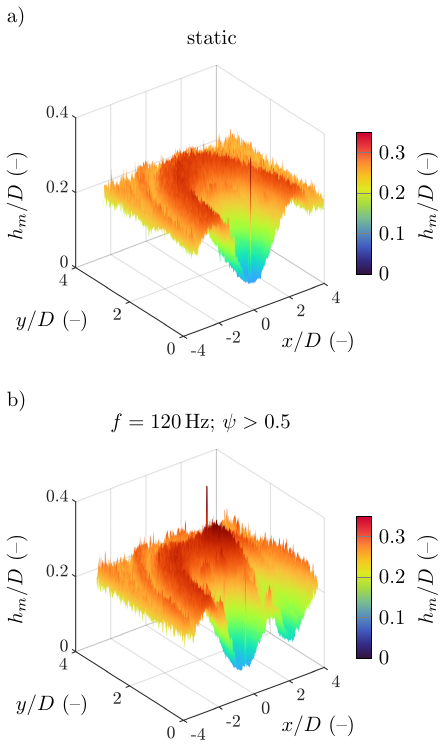}
	\caption{Cross section representation of 3D film height reconstruction for $\delta=0.22$, $We = 54$ and $Re = 3000$  obtained using two-color LIF at $t = 10~\mathrm{ms}$. \textbf{a)} static film. \textbf{b)} wavy film at $f= \SI{120}{\hertz}$, $d_s = 5~\mathrm{mm}$ and $\psi>0.5$.}
	\label{fig:11}
\end{figure}

For the static case, the reconstructed film topography shows the formation of a central cavity surrounded by an expanding rim, together with the propagation of capillary waves outward from the impact location. At higher impact energy ($We=110$, Fig. \ref{fig:10}c), the elevated rim reflects the formation of a rising liquid sheet and corona. When the impact occurs on a wavy film (Figs. \ref{fig:10}b,d), the 2C-LIF reconstruction reveals the strong influence of the underlying wave on droplet spreading. The spatial asymmetry of the rim and corona, previously observed qualitatively in Figs. \ref{fig:7}–\ref{fig:8}, is now quantitatively resolved. In both low and high Weber number cases, the portion of the rim or crown encountering the incoming wave exhibits increased height due to constructive superposition, whereas the section interacting with the pre-existing cavity generated by the acoustic wave source shows a noticeable depression.

\begin{figure*}
	\centering
	\includegraphics{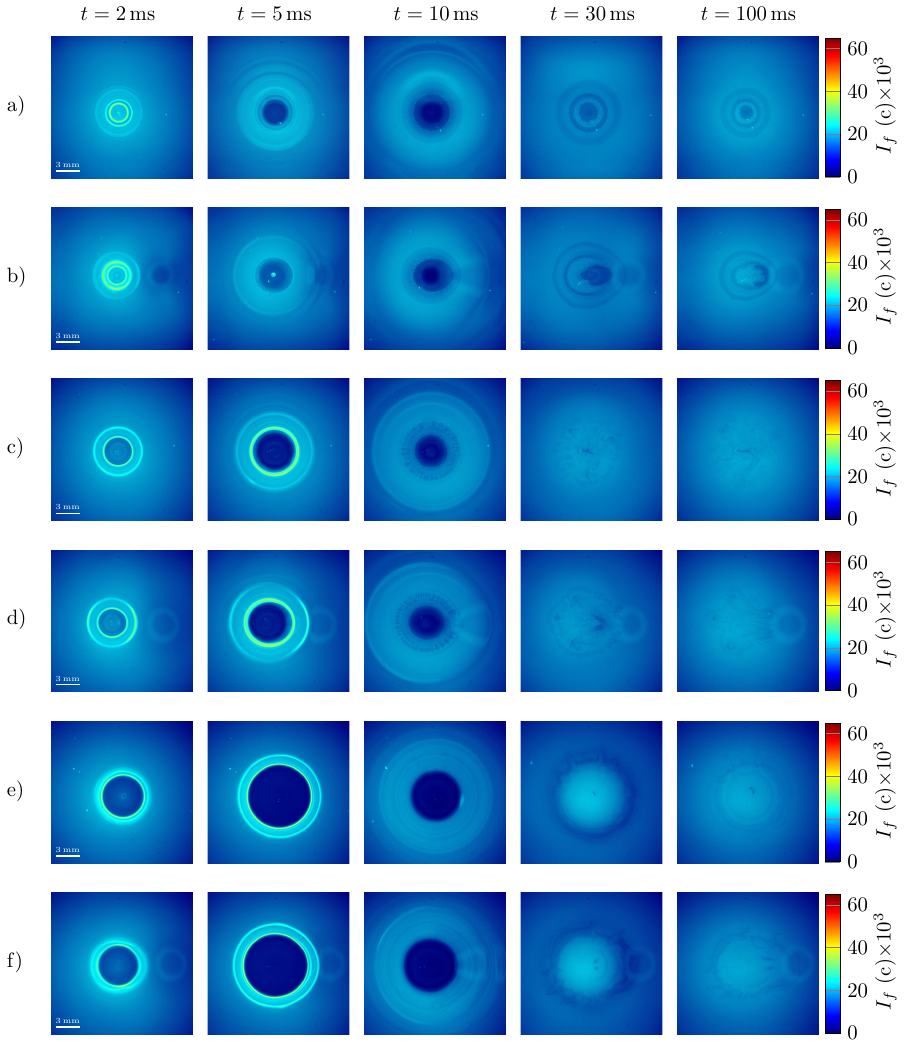}
	\caption{Bottom view raw images of fluorescent-labeled droplet impact on fluorescent-labeled thin liquid film of thickness $\delta=0.36$. \textbf{a)}, \textbf{c)} and \textbf{e)} Static films at $We=24$, $54$, $110$ and $Re=1800$, $3000$, $4200$, respectively. \textbf{b)}, \textbf{d)} and \textbf{f)} Wavy films generated at $f= \SI{120}{\hertz}$ with $d_s = 5~\mathrm{mm}$ and $\psi = 0$, for impact conditions $We=24$, $54$, $110$ and $Re=1800$, $3000$, $4200$, respectively. Colors represent the fluorescence signal intensity $I_{f}$.}
	\label{fig:12}
\end{figure*}

A more detailed comparison is shown in Fig. \ref{fig:11}, where the same $We=54$ case is displayed from a sectional perspective for both static film a) and the wavy film (b). The cross-sectional view emphasizes the difference in cavity geometry. This deformation pattern, which can only be observed from an oblique perspective, demonstrates the ability of the 2C-LIF method to capture local thickness variations even in highly dynamic impacts. Although the present study focuses primarily on species transport and concentrations distribution, the demonstrated capability to simultaneously resolve transient film topography establishes a strong basis for future work focused on quantifying the evolution of cavity depth and corona height during droplet impact on wavy liquid films.

\subsection{Species transport and mixing}
\label{sec:5.4}
Fig. \ref{fig:12} shows raw 2C-LIF images obtained for six representative cases at $\delta=0.36$, corresponding to three Weber numbers ($We=24$, $54$, and $110$) under two conditions: impact on a static film and impact on a wavy film generated at $f= \SI{120}{\hertz}$ with $d_s = 5~\mathrm{mm}$ and $\psi = 0$. The time sequence covers $t = 2$, $5$, $10$, $30$, and $100~\mathrm{ms}$.

\begin{figure*}[t]
	\centering
	\includegraphics[scale=0.77]{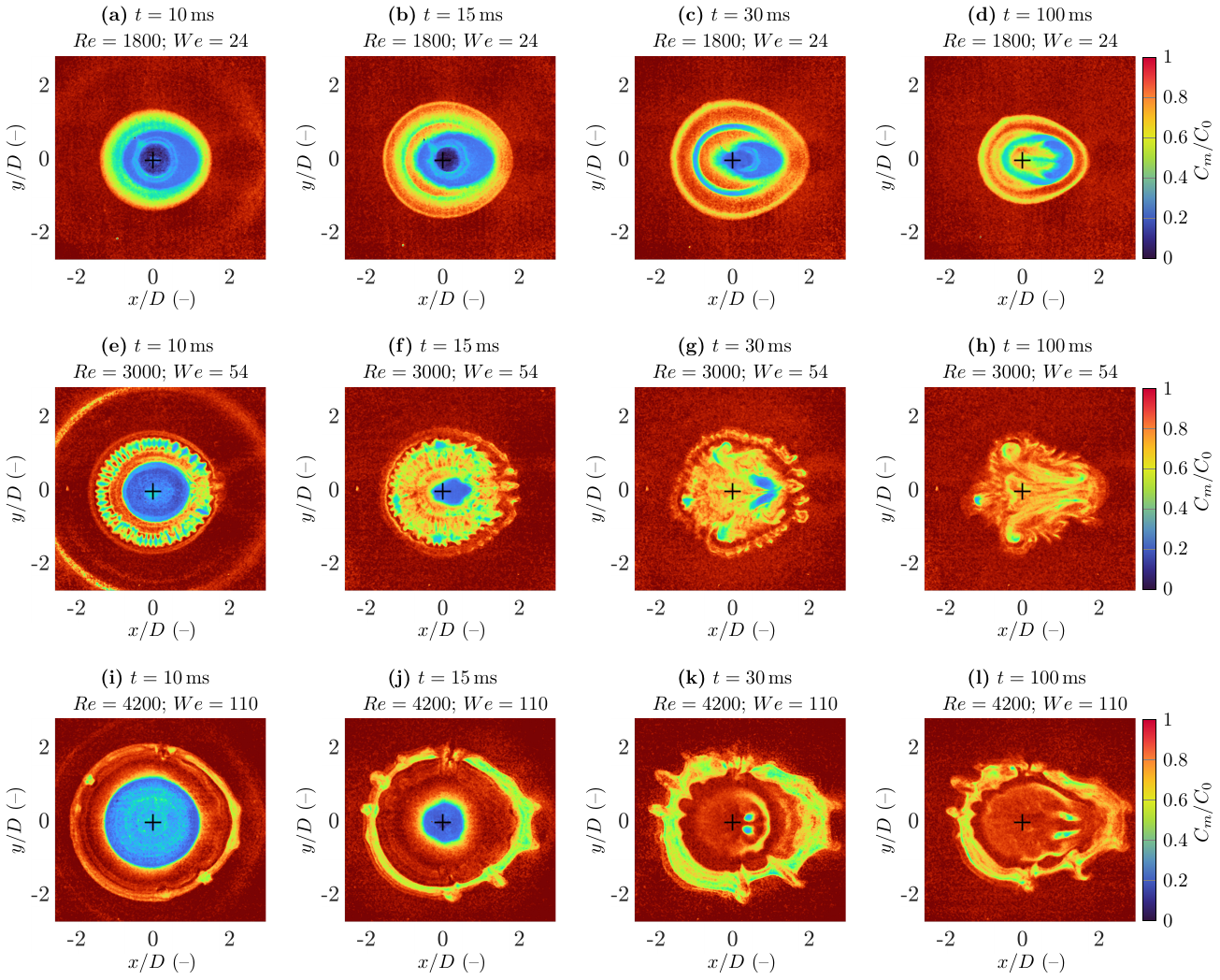}
	\caption{Temporal evolution of the normalized concentration field $C_m/C_0$ during droplet impact on a liquid film with $\delta=0.36$. Waves generated at $f= \SI{120}{\hertz}$ with $d_s = 5~\mathrm{mm}$ and $\psi = 0$. Rows corresponds to increasing impact velocity: \textbf{(a-d)} $Re=1800$, $We=24$; \textbf{(e-h)} $Re=3000$, $We=54$; \textbf{(i-l)} $Re=4200$, $We=110$. Columns show snapshots at $t=10$, $15$, $30$ and $=\SI{100}{\milli\second}$. The black cross marks the droplet impact location.}
	\label{fig:13}
\end{figure*}

During droplet impact on a liquid surface, strong velocity gradient arise at the liquid-liquid interface, which generate azimuthal vorticity. This vorticity rapidly rolls up into a vortex ring, as described by \citealt{cresswell1995drop}. For thin liquid films, as described in our previous work (\citealt{ennayar2025vortex}), interaction of this vortex ring with the underlying wall stretches it radially, causing boundary-layer separation and the formation of a secondary vortex rin that orbits around the first. When the impact energy is sufficiently high, these coupled vortex structures develop azimuthal instabilities that evolve into small-scale turbuluce (\citealt{walker1987impact,harris2012instability,cheng2010numerical,cerra1983experimental}). A stability map established in the work of \citealt{ennayar2025vortex} identified three mixing outcomes of droplet within thin liquid film during deposition regime depending on $Re$ and $\delta$. First one is Expansion of vortex ring for low Reynolds numbers, then multiple vortex ring formation and instability. In the present study, only the latter two regimes are investigated. The following measurements revisit these dynamics to serve as a reference for comparison with the wavy film impacts.

For $We=24$, $Re=1800$ static case (Fig. \ref{fig:12}a), the fluorescence intensity, represented by the colormap, increases shortly contact owing to local film thickening at the rim. At $t = 5~\mathrm{ms}$, a central crater develops, indicated by a darker region of reduced intensity, followed by the emergence of concentric ring structures at later times. These rings correspond to the primary vortex ring expanding radially along the wall and the formed secondary vortex ring due to boundary layer lift off. At $We=54$, $Re=3000$, illustrated in Fig. \ref{fig:12}c, azimuthal instabilities become visible as finger-like perturbations of the ring at $t = 10~\mathrm{ms}$. Such deformations arise from vortex rings' diameter compression during receding phase of the droplet impact (\citealt{ennayar2025vortex}). The effect of vortex-ring diameter compression on the onset of instability was also well documented by \citealt{saffman1979vortex} in their analysis of vortex-ring stability. These perturbations then lead to turbulent mixing patterns with a flower-like appearance by $t = 30~\mathrm{ms}$.

At higher Weber numbers ($We = 110$, $Re=4200$, Fig.~\ref{fig:12}e), the impact produces a well-defined central cavity that remains axisymmetric during the early stages. Unlike the $We=54$ case, no short-wavelength azimuthal perturbations are visible at $t = 10~\mathrm{ms}$. Instead, only a circular ring forms initially. At later times ($t = 30~\mathrm{ms}$), the ring develops long-wavelength distortions along parts of its circumference, indicating a different mixing regime. Previous study by \citealt{lee2015origin} on vortex ring formation during droplet impact on deep pools reported that for $We > 64$, the primary vortex ring is rapidly entrained upward by the rising jet. Additionally, the mixing regimes map in thin liquid film investigated by \citealt{ennayar2025vortex} was restricted to Weber numbers below 64. Hence, this regime boundary has not yet been examined for thin films impacts. The present results therefore suggest a possible transition in the dominant mixing mechanism for $We>64$, which warrants further investigation to elucidate the underlying vortex dynamics and their role in mixing at higher Weber numbers.

For impacts on wavy films at impact phase $\psi=0$ (Figs. \ref{fig:12}b,d,f), the fluorescence intensity changes reflects the effects of the pre-existing wave on the impact dynamics. At $We = 24$ and $Re=1800$, the fluorescence intensity at the impact center is higher compared to the static case, reflecting the larger local film thickness at the wave crest. Asymmetry develops in the spreading pattern, with the darker region in the impact location displaced toward the wave source. Moreover, micro-bubbles appear immediately after impact for both static and wavy films, consistent with the entrainment mechanism identified by \citet{thoroddsen2003air} for low Weber numbers, who showed that deformation of the free surface prior to contact traps a thick air sheet that breaks into bubbles upon collapse. For the cases $We=54$ (Fig. \ref{fig:12}d) and $We=54$ (Fig. \ref{fig:12}f), the influence of the wave remains evident. Rim spreading and internal flow structures become asymmetric, and the central liquid region is advected toward the wave source.

It should be noted that the circular bright feature visible on the right side of the images originates from reflections at the edge of the outlet tube, integrated into the acoustic wave generator. Its influence was accounted for during the calibration procedure, as additional reference measurements for the wavy film configuration were performed with the wave generator present in the optical field of the view at different $d_s$ distances.

Fig.~\ref{fig:13} presents the temporal evolution of the normalized concentration field $C_m/C_0$ obtained from 2C-LIF measurements for droplet impact on wavy thin liquid films at $\delta=0.36$. The waves are generated acoustically at an excitation frequency of $f= \SI{120}{\hertz}$, with the droplet impacting at a distance $d_s = 5~\mathrm{mm}$ from the wave source and at the crest of the wave ($\psi = 0$). Three impact conditions are shown corresponding to $Re=1800$, $3000$, and $4200$ ($We=24$, $54$, and $110$), respectively. The black cross marks the droplet impact location.

At $t = 10~\mathrm{ms}$, the concentration distributions already begin to deviate from the axisymmetric patterns observed for impacts on quiescent films. Although the same mixing mechanisms are initially present, the redistribution of droplet liquid gradually shifts toward the direction of the surface wave source, located on the right-hand side of the impact point. For $We=24$, the ring-shaped concentration structure, corresponding to the multiple vortex rings regime gets entrained toward the wave source. For $We=54$, where azimuthal perturbations develop due to vortex-ring instability, this displacement is less pronounced at these early times. A similar behavior is observed for $We=110$ in regards to concentration distribution in comparison to the quiescent case. 

At later times, the influence of the surface wave becomes increasingly pronounced. By $t = 30~\mathrm{ms}$, the droplet-rich region is clearly displaced relative to the droplet impact position, and the concentration field loses its axisymmetric structure. At $t = 100~\mathrm{ms}$, the asymmetry is fully developed for all cases. For $We=24$, the bulk of the droplet liquid is displaced toward the source of the acoustically generated wave, leading to a noticeable lateral shift of the low concentration region relative to the droplet impact location. Furthermore, the ring shaped vortex structures characteristic of the static case are distorted and elongated toward the acoustic source. 

\begin{figure*}
	\centering
	\includegraphics[scale=0.75]{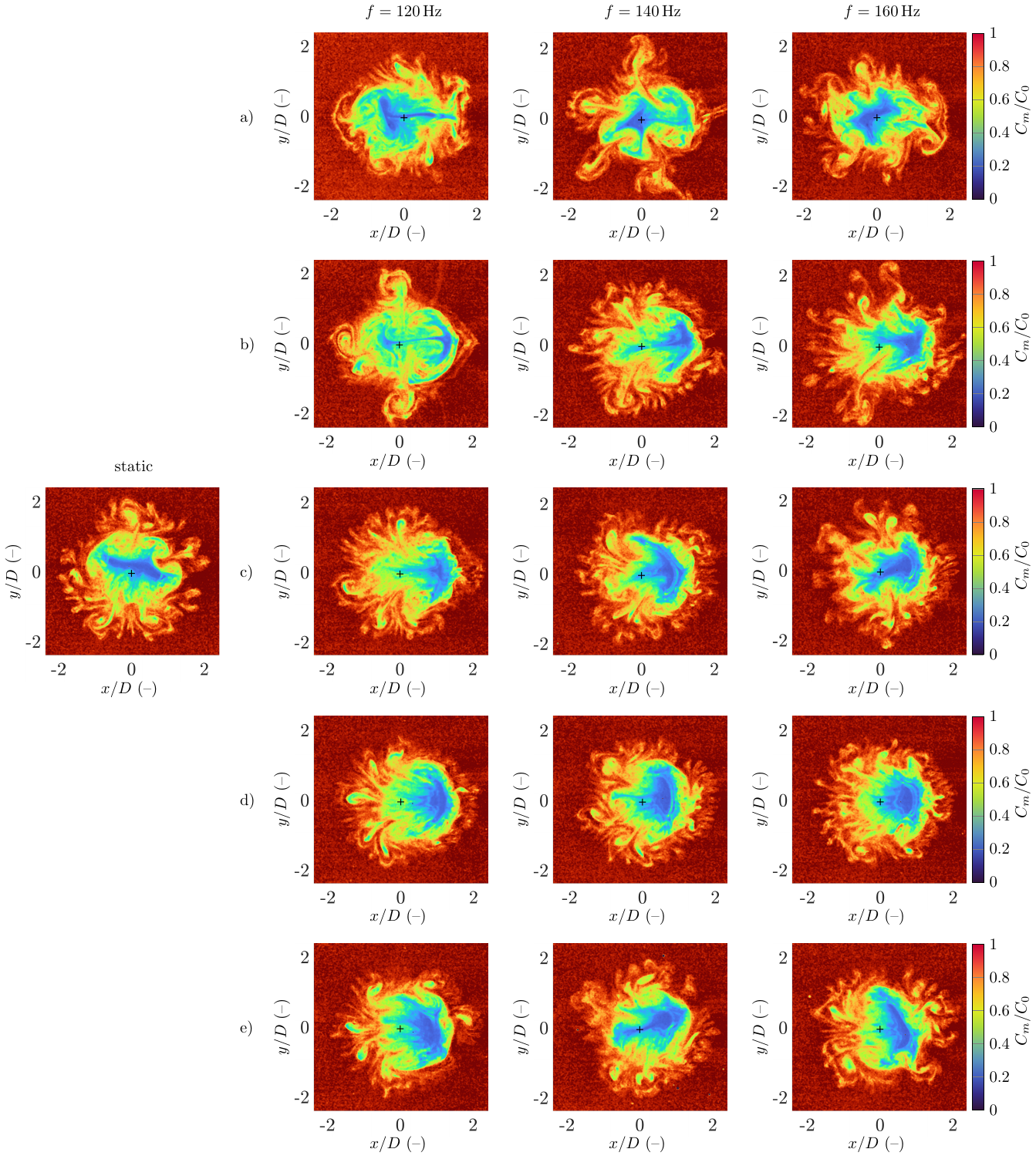}
	\caption{2D normalized concentration fields reconstructed using 2C-LIF for $\delta = 0.22$ at $t=100~\mathrm{ms}$ for droplet impact on wavy films at $We=54$, $Re=3000$ and $d_s = 5~\mathrm{mm}$.  \textbf{a)} $\psi>0.5$. \textbf{b)} $\psi\approx 0.25$. \textbf{c)} $\psi = 0$. \textbf{d)} $\psi\approx -0.25$. \textbf{e)} $\psi<-0.5$. The black cross marks the droplet impact location.}
	\label{fig:14}
\end{figure*}

For $We=54$, the instability-driven mixing remains dominant, yet a distinct redirection of the flow toward the surface wave source is observed. Additionally, the region that initially contained the highest droplet liquid fraction near the center becomes entrained toward the acoustic wave generator. Its concentration also sees an increase, indicating enhanced mixing with the liquid film during entrainment. A comparable behavior occurs at $We=110$, with the droplet-rich zone again drawn toward the wave source, confirming that the surface wave introduced by the acoustic excitation drives preferential redirection of the flow as well as disrupts the axial symmetry characteristic of impacts on quiescent films.

The influence of surface wave parameters on droplet mixing can be examined closely by impacts performed at different excitation frequencies, phases as well as different $d_s$ values. Fig. \ref{fig:14} displays concentration distributions at $t = 100~\mathrm{ms}$ for $\delta = 0.22$, $We = 54$, $Re = 3000$, and $d_s = 5~\mathrm{mm}$. The dataset includes one static reference case and fifteen impacts on wavy films, corresponding to three excitation frequencies ($f=120$, $140$ and $\SI{160}{\hertz}$) and five impact phases: (a)~pre-front ($\psi > 0.5$), (b)~front-slope ($\psi \approx 0.25$), (c)~crest ($\psi = 0$), (d)~back-slope ($\psi \approx -0.25$), and (e)~far-back ($\psi < -0.5$). All configurations exhibit vortex-ring instability leading to turbulent mixing.

A key observation from these data is that the liquid near the center of impact is displaced toward or away from the acoustic wave generator depending on $\psi$. For $f=\SI{120}{\hertz}$, the central region in the ''pre-front" case is driven away from the wave generator. As $\psi$ decreases toward the ''front-slope" regime, the central liquid fraction is split, where parts moves away from , and parts toward, the wave source. For impacts near the crest, back-slope, and far-back positions ($\psi \approx 0$, $\psi \approx -0.25$, and $\psi < -0.5$), the entire droplet-rich region is directed toward the generator, consistent with the redirection observed in Figs.~\ref{fig:12} and \ref{fig:13}. Increasing the excitation frequency, and thus reducing the wave amplitude, weakens this displacement. For $\psi > 0.5$ the outward displacement becomes less pronounced , while for $\psi \approx 0.25$ and remaining phases, the flow remains oriented toward the wave generator but with shorter displacement distances. 

\begin{figure}
	\centering
	\includegraphics[scale=0.85]{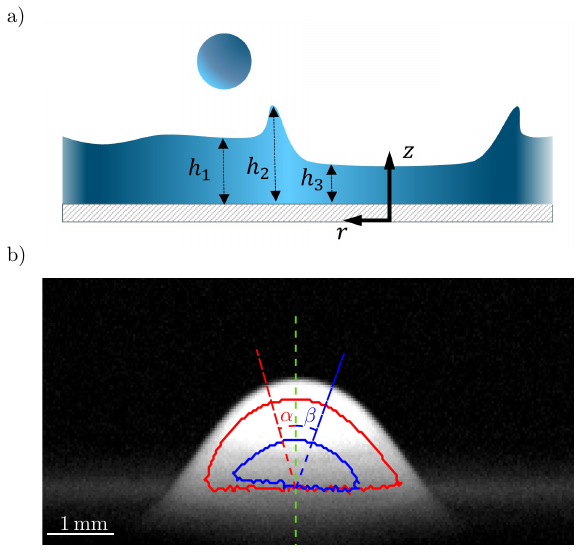}
	\caption{\textbf{a)} Schematic representation of the local thickness variation across a wavy thin liquid film. \textbf{b)} inclined jet formation during droplet impact on wavy thin liquid and compared to static film. (red) $\psi > 0.5$ and (blue) $\psi<-0.5$.}
	\label{fig:15}
\end{figure}

To elucidate the origin of this behavior, the influence of local film depth prior to impact is considered. Fig. \ref{fig:15}a schematically illustrates the variation in film thickness across the wave profile. $h_1$ denotes the depth in the pre-front region, approximately equal to the initial quiescent thickness $h$. $h_2$ corresponds to the maximum depth at the crest and $h_3$ represents the reduced depth within the acoustically-induced cavity near the far back region. This variation in local film thickness establishes a depth gradient across the impact area, which in turn alters the subsequent droplet-induced cavity retraction dynamics. In fact, an analogous mechanism was reported by \citet{lee2020symmetry}, who demonstrated that when a droplet impacts a pool with a depth gradient created by a sloped bottom, asymmetric cavity collapse occurs. The difference in local depth leads to variations in the propagation speed of the capillary waves generated by the droplet impact, which is given by he capillary wave dispersion relation \cite{currie2002fundamental}:
\begin{equation}
	c \sim \sqrt{\frac{\sigma k}{\rho}  \tanh(kh)},
	\label{eq:3}
\end{equation}
where $k$ is the wavenumber. Since $\tanh(kh)$ increases with $h$, capillary waves propagates faster on the deeper side. Consequently, the cavity apex shifts toward the shallower side during collapse, producing a jet tilted in the same direction as the region of reduced depth.

\begin{figure*}[t]
\centering
\includegraphics{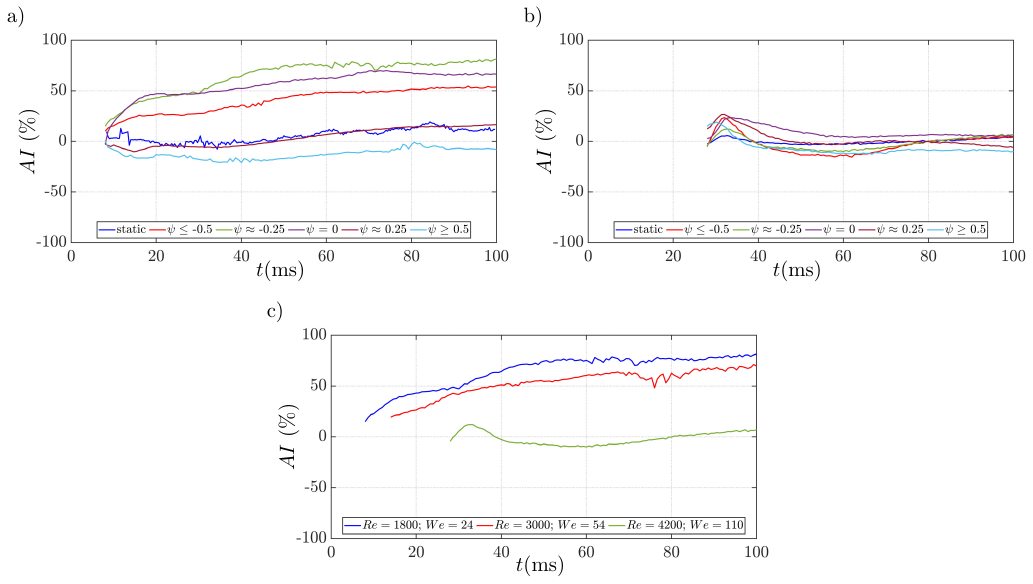}
  \caption{The temporal variation of the asymmetry index $AI$ for impacts on film of thickness $\delta=0.36$ and waves generated with frequency $f=\SI{120}{\hertz}$ at distance $d_s = 5~\mathrm{mm}$. \textbf{a)} $We=24$, $Re=1800$. \textbf{b)} $We=110$, $Re=4200$. \textbf{c)} Comparison between different Weber number for impact phase $\psi\approx -0.25$.}
    \label{fig:18}
\end{figure*}


The same mechanism explains the present observations. In the pre-front case,  the right-hand side of the cavity encounters the traveling surface wave, where the local depth exceeds that on the opposite, resulting in asymmetric collapse toward the thinner region (i.e., away from the wave source). Conversely, for impacts on or beyond the crest, part of the cavity forms over a local depression, driving the retraction and the liquid near the central area toward the wave generator. Experimental evidence supporting this interpretation is shown in Fig. \ref{fig:15}b, comparing static, pre-front and far-back impacts for $\delta = 0.36$, $We = 110$, and $Re = 4200$. The pre-front case corresponds to $f=\SI{120}{\hertz}$ and $d_s = 5~\mathrm{mm}$, while the far-back case was obtained at $f=\SI{140}{\hertz}$ and $d_s = 7.5~\mathrm{mm}$. The latter configuration was chosen to clearly visualize the jet formation during impact. While the static case yields a vertical Worthington jet, the pre-front case shows a slight tilt toward the thinner region, corresponding to the region in front of the surface wave for this case. On the other hand, the far-back case produces a jet inclined toward the acoustic wave generator. This jet inclination confirms that cavity asymmetry, governed by local film depth gradients, dictates the redirection of the flow during droplet impact on wavy films.

\begin{figure}
	\centering
	\includegraphics[scale=0.925]{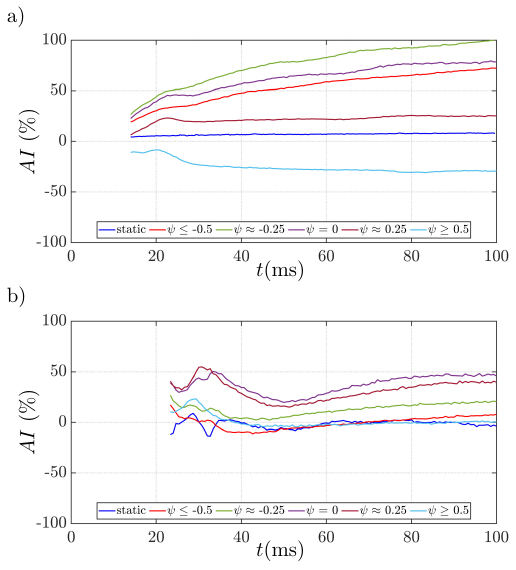}
	\caption{The temporal variation of the asymmetry index $AI$ for impacts on film of thickness $\delta=0.22$ and waves generated with frequency $f=\SI{120}{\hertz}$ at distance $d_s = 5~\mathrm{mm}$. \textbf{a)} $We=54$, $Re=3000$. \textbf{b)} $We=110$, $Re=4200$.}
	\label{fig:19}
\end{figure}

\subsection{Asymmetry index}
\label{sec:5.5}
To quantitatively assess the directional influence of surface waves on droplet mixing, a dimensionless asymmetry index ($AI$) was introduced. The aim of this index is to measure how the imposed wave modifies the left--right symmetry of the concentration field. Throughout this study, the right side corresponds to the side facing the wave generator.

The analysis was performed within a region of interest defined based on the maximum lateral expansion during the impact as well as the detectable concentration  gradient between the actively mixed region and the surrounding undisturbed film areas retaining their initial concentration. This region was divided into two equal halves relative to the impact center. For each half, the coefficient of variation was calculated separately, giving $CV_R$ for the right side and $CV_L$ for the left side. The coefficient of variation is defined as the standard deviation of the local concentration normalized by its spatial mean,
\begin{equation}
	CV(t) = \frac{\sqrt{\frac{1}{n-1}\displaystyle\sum_{x,y} \Bigr[C_{m}(x,y,t)-\frac{1}{n}\sum_{x,y} C_{m}(x,y,t)\Bigr]^2}}{\frac{1}{n}\displaystyle\sum_{x,y} C_{m}(x,y,t)},
	\label{eq:4}
\end{equation}
where $n$ denotes the number of pixels within the chosen mixing region and $C_m(x,y,t)$ is the measured local concentration at time $t$. A lower $CV$ indicates a more homogeneous concentration distribution, whereas a higher $CV$ indicates stronger spatial concentration variations.

In an ideally symmetric mixing process, the two sides of the impact region would yield comparable values of $CV_R$ and $CV_L$. The presence of surface waves, however, can modify the flow and concentration fields, producing stronger concentration variations on one side than on the other. Therefore, $AI$ quantifies the directional imbalance in mixing and identifies which side exhibits the larger concentration variability.

To achieve this, a fold-change based equation was adopted. Unlike subtraction based or normalized difference formulas, this approach is symmetric and scale-invariant, allowing the results to be directly interpreted in terms of how many times greater the variability in on one side compared with the other. The asymmetry index is expressed as:
\begin{equation}
	AI = \mathrm{sgn}\!\left(\ln\frac{CV_R}{CV_L}\right)
	\left[\exp\!\left(\left|\ln\frac{CV_R}{CV_L}\right|\right) - 1\right] \times 100,
	\label{eq:5}
\end{equation}

\begin{figure*}
	\centering
	\includegraphics{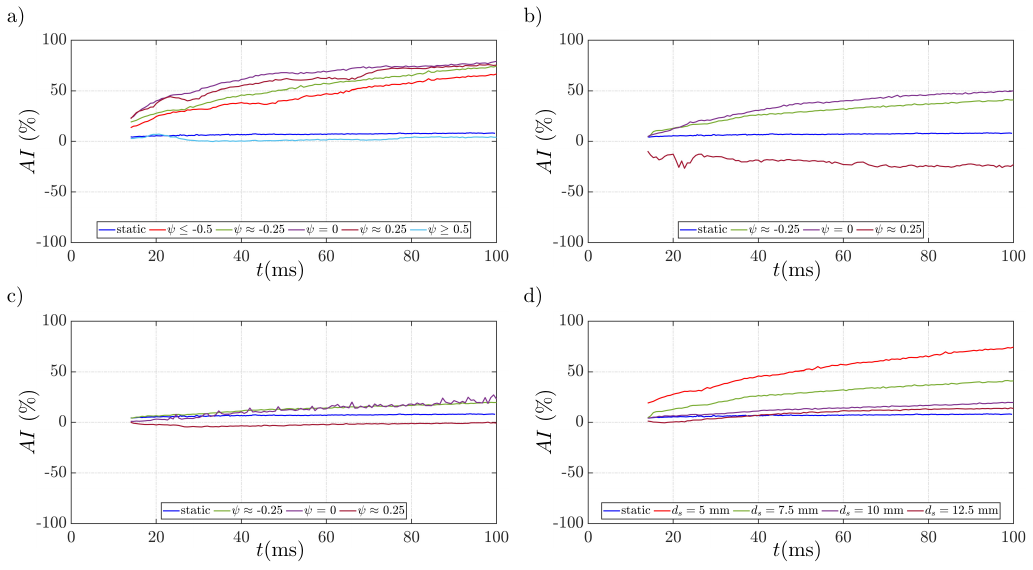}
	\caption{The temporal variation of the asymmetry index $AI$ for impacts on film of thickness $\delta=0.22$ at $We=54$, $Re=3000$ and waves generated with frequency $f=\SI{40}{\hertz}$. \textbf{a)} $d_s = 5~\mathrm{mm}$. \textbf{ab)} $d_s = 7.5~\mathrm{mm}$. \textbf{c)} \textbf{c)} Comparison between different distances  for impact phase $\psi\approx -0.25$.}
	\label{fig:20}
\end{figure*}

Positive $AI$ values indicate that variability is greater on the right-hand side toward the wave generator, whereas negative values signify dominance on the left-hand side. For instance, $AI = +50\%$ implies that the right side exhibits $1.5\times$ higher variability than the left, while $AI = -50\%$ indicates the inverse.

The temporal evolution of the asymmetry index ($AI$) for $\delta = 0.36$ is shown in Fig. \ref{fig:18}. The dataset corresponds to impacts performed at frequency of $f=\SI{120}{\hertz}$ and a distance of $d_s = 5~\mathrm{mm}$, covering various impact phases. In Fig. \ref{fig:18}a, corresponding to $We=24$ and $Re=1800$, the static case remains centered near $AI=0$, confirming symmetric mixing between both sides. When surface waves are present, however, the curves show clear phase-dependent deviations. Impacts on the crest, back-slope, and far-back regions yield positive $AI$ values, indicating greater variability on the right side toward the wave generator and consistent with the findings discussed earlier. In contrast, for the pre-front case ($\psi>0.5$), the left side dominates, yielding negative $AI$ values, whereas the front-slope case slightly favors the left side initially before gradually converging to similar value as quiescent films. 

At higher impact energy ($We = 110$, Fig. \ref{fig:18}b), the amplitude of the asymmetry decreases substantially. The $AI$ values for all phases remain close to those of the static film, demonstrating that strong inertial mixing tend to homogenize the concentration field thereby masking wave-induced directional effects. This trend is further confirmed in Fig. \ref{fig:18}c, where for $\psi \approx -0.25$, the values of $AI$ gradually converge toward zero with increasing $We$.

For the thinner film ($\delta=0.22$), Fig. \ref{fig:19} presents analogous results. In Fig.~\ref{fig:19}a ($We = 54$, $f = \SI{120}{\hertz}$, $d_s = 5~\mathrm{mm}$), the phase dependence again governs the sign and magnitude of asymmetry.
The pre-front case produces negative $AI$, confirming that the droplet-induced cavity retracts toward the left, while all other phases show positive values corresponding to right-directed motion. When the impact velocity increases ($We = 110$, Fig.~\ref{fig:19}b), $|AI|$ values become smaller overall, and the far-back and pre-front cases nearly overlap with the static reference, indicating a reduced sensitivity to the influence of the surface wave.

The influence of the wave generator distance to the droplet impact location $d_s$ is examined for $\delta = 0.22$, $We = 54$, $Re = 3000$, and $f = \SI{140}{\hertz}$. Figs. \ref{fig:20}a–c show $AI(t)$ for increasing distances ($d_s = 5$, $7.5$, and $10~\mathrm{mm}$), while Fig. \ref{fig:20}d summarizes the trends for the back-slope phase. At $d_s = 5~\mathrm{mm}$, asymmetry remains clearly positive for most phases, with the right-hand side exhibiting stronger variability. For the pre-front case, due to the smaller wave amplitude resulting from the higher excitation frequency ($f = \SI{140}{\hertz}$), $AI$ remains close to zero. This behavior is supposed to result from the influence of the acoustically generated cavity, which counteracts the leftward displacement of the droplet liquid and produces an approximate balance between both sides of the mixing region. Consequently, the resulting asymmetry stays close to the static reference condition.

As $d_s$ increases, the amplitude of $|AI|$ systematically decreases, and all curves progressively converge toward the static reference, showing that the effect of the surface wave on mixing weakens with distance from the source. This observation highlights that the primary driver of asymmetry is the cavity generated by the acoustic excitation. The cavity introduces a pronounced local depth gradient that strongly alters the retraction dynamics of the droplet cavity when the impact occurs nearby. However, as the droplet impacts farther from the wave generator, this gradient becomes progressively smaller, and the influence of the acoustically-induced cavity diminishes. The resulting film behaves increasingly like a quiescent layer and the $AI$ approaches zero.

Furthermore, the front-slope case provides a clear illustration of the influence of the cavity. At larger $d_s$, it exhibits negative $AI$ values, showing that the collapse and displacement shifts away from the wave generator toward the left once the acoustically-induced cavity no longer shapes the local surface profile. This behavior signals the transition from a cavity-dominated regime to a capillary-wave regime, where  the local slope of the surface at the impact location determines the direction of the asymmetry. In fact, as seen for the pre-front case in Fig. \ref{fig:20}a, the interaction between these two regimes can locally balance out, leading to nearly symmetric mixing conditions.

Overall, the $AI$ analysis confirms that the asymmetry in droplet mixing is predominantly governed by the localized depth variations created by the acoustical wave generator. As the impact location shifts farther from the source, the cavity effect weakens and the regime becomes dominated by the capillary wave itself, where the surface slope at the impact point dictates the remaining directionality of the flow. Beyond this region, where only small-amplitude capillary waves persist, the mixing dynamics resemble those of a quiescent film, with negligible asymmetry.

\section{Conclusion}
\label{sec:6}
The present work demonstrated how traveling surface waves alter the impact dynamics and mixing behavior of droplet droplet impinging on thin liquid films. By generating controlled capillary waves using an acoustic excitation system, wave-induced topographies analogous to those produced by preceding droplets were reproduced without introducing additional mass or concentration disturbances into the film. 

Droplet impacts on wavy films exhibited significant deviations from impacts on quiescent films. The rim, cavity and crown evolution became strongly asymmetric and jet formation was suppressed when the droplet encountered sufficiently large wave amplitudes. These effects were governed by the phase of the wave relative to impact. The directional bias in the flow was traced to local variation in film depth, which modify cavity collapse similar to findings by \citet{lee2020symmetry} in their work regarding symmetry breaking jets by gradients in liquid pool depth. Impacts on regions of increased depth shifted the cavity apex toward the shallower side, while impacts on or behind the crest led to cavity collapse toward the wave source. These mechanisms were confirmed by the inclination of the resulting jets and by the deformation of the near-field flow.

The influence of the surface wave extended to the mixing process. Depth-averaged concentration fields showed that the droplet-rich liquid part located at the center is displaced toward or away from the wave generator depending on the local depth profile at the impact point. At moderate Weber numbers, vortex ring instability and subsequent turbulent mixing remained present but developed a pronounced asymmetry. Increasing the wave amplitude intensified this displacement, while shifting the impact location farther from the acoustic wave generator reduced the effect. The asymmetry index introduced in this work quantified this asymmetry and revealed a clear transition from a cavity-dominated regime near the excitation site to a regime where only the surface slope governs the flow redirection. At sufficiently high Weber numbers, inertial mixing attenuated these asymmetries, and the mixing dynamics approached those of quiescent films.

Overall, the results show that even small-to-moderate amplitude surface waves can have a measurable influence on droplet impact and mixing in thin films. The combination of controlled wave generation and simultaneous thickness and concentration measurement offers a framework for isolating and quantifying these effects. The findings indicate that surface disturbances generated by prior droplets or external forcing cannot be neglected when evaluating droplet–film interactions in applications involving repeated or closely spaced impacts.

\section*{Declarations}
\textbf{Ethical approval} This declaration is not applicable.\\\\
\textbf{Competing interests} The authors have no competing interests to declare that are relevant to the content of this article.\\\\
\textbf{Funding} This project is funded by the Deutsche Forschungsgemeinschaft (DFG, German Research Foundation) – project number 237267381 – TRR 150, sub-project A07.\\\\



\bibliographystyle{apalike}
\bibliography{Bibliography_LIF_Drop}

\appendix       

\end{document}